\documentclass[sigconf]{acmart}
\setcopyright{none}
\renewcommand{\footnotetextcopyrightpermission}[1]{}
\acmConference{}{}{}
\acmISBN{}
\acmDOI{}

\usepackage{tikz}
\usepackage{amsmath}
\usepackage{amssymb}
\usepackage{amsthm}
\usepackage{pifont}

\usepackage{enumitem}
\usepackage{tcolorbox}

\usepackage{multirow}
\usepackage{graphicx}
\usepackage{subcaption}
\usepackage{xcolor}
\usepackage{tabularx}
\usepackage[ruled,vlined]{algorithm2e}
\tcbuselibrary{listings,breakable}

\newcommand{\nop}[1]{}
\newtheorem{theorem}{Theorem}

\newtheorem{definition}[theorem]{Definition}

\newenvironment{ditemize}{
\begin{list}{{\bf $\bullet$}}{
\setlength{\itemsep}{0pt}\setlength{\rightmargin}{0pt}
\setlength{\leftmargin}{1.2em}\setlength{\parsep}{0pt}}}{
\end{list}}

\newcommand{\white}[1]{
  \begingroup
  \setlength{\fboxsep}{0pt}
  \colorbox{white}
  \endgroup
}
\newcommand{\system}{\textsf{RelGuard}}

\newcommand{\ts}{s_{\text{target}}}

\begin{document}

\title{Differentially Private GraphRAG: Protecting Relation Connectivity and Semantics}

\author{Jie Fu}
\affiliation{
  \institution{Stevens Institute of Technology}
  \city{Hoboken}
  \state{NJ}
  \country{USA}
}
\email{jfu13@stevens.edu}

\author{Yuan Hong}
\affiliation{
  \institution{University of Connecticut}
  \city{Storrs}
  \state{CT}
  \country{USA}
}
\email{yuan.hong@uconn.edu}

\author{Yue Ning}
\affiliation{
  \institution{Stevens Institute of Technology}
  \city{Hoboken}
  \state{NJ}
  \country{USA}
}
\email{Yue.Ning@stevens.edu}

\author{Wendy Hui Wang}
\affiliation{
  \institution{Stevens Institute of Technology}
  \city{Hoboken}
  \state{NJ}
  \country{USA}
}
\email{hwang4@stevens.edu}

\renewcommand{\shortauthors}{Fu et al.}

\begin{abstract}
Graph Retrieval-Augmented Generation (GraphRAG) connects information across documents through entity relationships to support relational reasoning. However, black-box access to GraphRAG systems can enable adversaries to reconstruct knowledge graphs and recover both structural connectivity between entities and the semantics of their relationships. Existing defenses lack formal privacy guarantees and provide limited protection against such reconstruction attacks. We propose \system, the first differentially private framework for GraphRAG, which provides formal privacy guarantees by mitigating connectivity leakage and protecting relations and their semantic meaning through relation removal.  We formalize  {\em relation differential privacy}  (relation-DP), which bounds the distinguishability of mechanism outputs according to the semantic distance between alternative relations while accounting for relation removal. Building on this definition, we develop a four-step defense mechanism that sanitizes source documents before GraphRAG indexing by perturbing relation predicates or removing their associated connectivity using the exponential mechanism. Experiments across three GraphRAG systems and two targeted reconstruction attacks demonstrate that \system\ provides stronger protection against reconstruction attacks than four existing defense mechanisms while better preserving downstream  utility.
\end{abstract}


\maketitle
\pagestyle{plain}
\thispagestyle{plain}

\section{Introduction}

Graph Retrieval-Augmented Generation (GraphRAG) integrates knowledge graphs, which represent entities and their relationships, with unstructured text corpora to support knowledge retrieval, domain-specific question answering, and multi-hop reasoning in Web services \cite{edge2024local,peng2025graph,guo2025lightrag}. By modeling relationships among entities, GraphRAG enables information scattered across documents to be connected and retrieved through graph structures, enhancing the relational reasoning capabilities of large language models (LLMs).

However, the relational information that enables these capabilities also creates privacy risks. We consider two complementary dimensions of relation privacy: (1) {\bf structural connectivity}, which reveals whether two entities are connected in the knowledge graph, and (2) {\bf relation semantics}, which reveals the nature of the relationship between connected entities. An adversary who infers connectivity can discover previously unknown associations between entities, while recovering relation semantics can further reveal the specific nature of those associations. Recent studies have demonstrated that adversaries can recover such information by reconstructing knowledge graphs through black-box access to GraphRAG systems \cite{liu2026exposing,yang2026query,gu2026graphsteal,song2026graphs}. Although existing defense mechanisms attempt to mitigate these reconstruction attacks \cite{gu2026graphsteal,song2026graphs,zeng2024good}, they lack formal privacy guarantees, leaving the extent of information leakage difficult to characterize and control.

To address this limitation, we propose \system, the first differentially private framework for  GraphRAG, protecting relation semantics while mitigating connectivity leakage. Our key insight is that relation protection must account for both the existence of a relationship and its meaning when present. Simply suppressing edges can protect connectivity but may unnecessarily impair downstream reasoning, whereas replacing relation predicates without accounting for edge removal does not provide a unified treatment of both privacy dimensions.

First, we formalize {\em relation differential privacy (relation-DP)}, a metric-based privacy definition that calibrates output indistinguishability to the semantic distance between alternative relation predicates while also accounting for the possibility of relation removal. This formulation bounds the distinguishability of released relation semantics, while relation removal helps protect connectivity.

Second, we develop a four-step defense mechanism that sanitizes private documents before they are indexed by GraphRAG. Specifically, \system: (1) extracts canonical entity--relation triples from the private documents; (2) generates candidate relation predicates for each extracted relation; (3) perturbs each relation by sampling a replacement predicate or a \texttt{null} outcome, which suppresses the corresponding connectivity; and (4) reconstructs sanitized documents by consistently updating relation expressions or removing sentences associated with \texttt{null} outcomes. GraphRAG then builds its index from the sanitized documents and uses the resulting index for query answering. This design integrates formal relation privacy protection into the GraphRAG pipeline while seeking to preserve the information needed for accurate question answering.

We evaluate \system\ on three GraphRAG systems and two real-world datasets, comparing it against four existing defense mechanisms. \system\ achieves the lowest relation recovery F1 across all the settings and the lowest connection recovery F1 in most settings. In particular, it reduces relation recovery F1 by 26.98\%--44.35\% compared with the strongest baseline. Meanwhile, \system\ achieves the best privacy--utility trade-off, limiting the reduction in question-answering F1 to just 2.98\%--8.97\%. These results demonstrate that \system\ provides stronger protection against relation reconstruction while preserving high utility for downstream question answering.

In summary, we make the following contributions:

\begin{ditemize}
    \item We introduce \emph{relation differential privacy (relation-DP)}, a metric-based privacy framework for formally protecting relation semantics and mitigating connectivity leakage in GraphRAG.

    \item We design a relation-DP-based defense mechanism that sanitizes source documents before GraphRAG indexing, jointly protecting relation semantics and connectivity through relation substitution and removal.

    \item Our extensive empirical evaluation demonstrates the advantages of \system\ over existing defenses in terms of both privacy protection and utility preservation.
\end{ditemize}

\section{Background}
\label{sec:preliminaries}

\subsection{Graph Retrieval-Augmented Generation}
\label{sec:prelim_graphrag}

{\em Graph Retrieval-Augmented Generation} (GraphRAG) uses graph-structured
knowledge to retrieve connected evidence for LLM generation
\cite{edge2024local,peng2025graph,guo2025lightrag}.
Systems either retrieve from existing knowledge graphs or construct
graph indexes from documents~\cite{peng2025graph}.
We focus on document-based GraphRAG, which extracts entities and
relations from source text and retains associated passages as evidence.

\noindent\textbf{Index construction.}
Given a private document $D$, the system constructs
an index $\mathcal I=\mathsf{Index}(D)$ containing a graph
$G=(V,E)$. Each triple $(s,r,o)$ represents a relation $r$ from
subject entity $s$ to object entity $o$. The index may also store
source passages, entity and relation descriptions, community summaries,
and vector representations
\cite{edge2024local,guo2025lightrag}.

\noindent\textbf{Retrieval and generation.}
For a query $q$, the retriever selects evidence
$C_q=\mathsf{Retrieve}(q,\mathcal I)$, which the LLM uses to generate
an answer $\widehat y=\mathsf{Generate}(q,C_q)$.
LightRAG uses dual-level retrieval~\cite{guo2025lightrag}, PathRAG
retrieves relational paths~\cite{chen2026pathrag}, and HippoRAG 2
uses associative retrieval~\cite{gutierrez2025rag}.
Retrieved evidence can include both graph relationships and associated text.

\subsection{Differential Privacy}
\label{sec:prelim_dp}

Differential privacy (DP) bounds how much the output distribution of a
randomized mechanism can change when its protected input changes
\cite{dwork2006calibrating,dwork2014algorithmic}.

\begin{definition}[$\epsilon$-Differential Privacy]
\label{def:background_dp}
Let $x\sim x'$ denote neighboring inputs differing in one protected
unit. A randomized mechanism $\mathcal M:\mathcal X\rightarrow
\mathcal Y$ satisfies $\epsilon$-DP if, for every neighboring pair
and every measurable output event $\mathcal S\subseteq\mathcal Y$,
\begin{equation}
\Pr[\mathcal M(x)\in\mathcal S]
\leq e^\epsilon\Pr[\mathcal M(x')\in\mathcal S].
\label{eq:background_dp}
\end{equation}
\end{definition}

Smaller $\epsilon$ imposes a tighter bound on distinguishability.
Standard DP specifies protected comparisons through binary adjacency.
Metric-DP instead uses an input distance to express how the privacy
bound varies across possible values
\cite{chatzikokolakis2013broadening,fernandes2019generalised}.

\begin{definition}[$\epsilon$-Metric Differential Privacy]
\label{def:background_metric_dp}
For a metric space $(\mathcal X,d_{\mathcal X})$, a randomized
mechanism $\mathcal M$ satisfies $\epsilon$-metric-DP if, for every
$x,x'\in\mathcal X$ and measurable output event $\mathcal S$,
\begin{equation}
\Pr[\mathcal M(x)\in\mathcal S]
\leq e^{\epsilon d_{\mathcal X}(x,x')}
\Pr[\mathcal M(x')\in\mathcal S].
\label{eq:background_metric_dp}
\end{equation}
\end{definition}

The input distance $d_{\mathcal X}$ in $\epsilon$-metric-DP must satisfy the triangle
inequality: $d_{\mathcal X}(x,z)\leq d_{\mathcal X}(x,x')
+d_{\mathcal X}(x',z)$ for all $x,x',z\in\mathcal X$. Nearby inputs receive a tighter bound; more distant inputs permit
larger differences between output distributions.
Metric-DP captures semantic differences through input distances.
Text-privacy methods use Euclidean distance between word
embeddings~\cite{feyisetan2020privacy}, Mahalanobis distance in
sentence-embedding space~\cite{bollegala2025metric}, and Earth
Mover's distance between bags of word
embeddings~\cite{fernandes2019generalised}.
The metric and its scale must be specified together with $\epsilon$.

A common approach to achieving $\epsilon$-metric-DP is the
\emph{exponential mechanism}, which selects from a finite candidate
set according to a utility score~\cite{mcsherry2007mechanism,chatzikokolakis2013broadening}.
Let $x\in\mathcal X$ be the private input, $\mathcal Y$ a fixed finite
output set, and $u:\mathcal X\times\mathcal Y\to\mathbb R$ a utility
function satisfying
\[
|u(x,y)-u(x',y)|\leq d_{\mathcal X}(x,x')
\]
for all $x,x'\in\mathcal X$ and $y\in\mathcal Y$.
Sampling according to
\begin{equation}
\Pr[\mathcal M_{\mathrm{EM}}(x)=y]
=\frac{\exp(\epsilon u(x,y)/2)}
{\sum_{z\in\mathcal Y}\exp(\epsilon u(x,z)/2)}
\label{thm:exponential_mechanism}
\end{equation}
satisfies $\epsilon$-metric-DP. Higher-utility candidates are more
likely, with $\epsilon$ controlling concentration.

\section{Related Work}
\label{sec:related_work}

\noindent\textbf{Privacy inference in GraphRAG.}
RAG systems are vulnerable to text
extraction~\cite{cohen2024unleashing,jiang2024rag} and membership
inference~\cite{li2025budgetleak}. GraphRAG attacks further recover
entity connections and relation semantics from generated answers.
Untargeted attacks seek broad graph recovery: AGEA uses graph memory
and novelty-guided exploration~\cite{yang2026query}, while GraphSteal
expands entity neighborhoods through breadth-wise
traversal~\cite{gu2026graphsteal}. Targeted attacks recover relations
around a specified entity. GraphSteal uses depth-wise search, whereas
GRASP frames extraction as structured relation processing with record
identifiers and adaptive queries~\cite{song2026graphs}. We evaluate
both connection and semantic relation recovery under targeted attacks.

\noindent\textbf{Privacy defenses for GraphRAG.}
Existing defenses modify generation instructions or retrieved
contexts. Protective Prompt restricts disclosure of retrieved
content~\cite{gu2026graphsteal}, while Abstractive requires
query-relevant paraphrasing~\cite{liu2026exposing,song2026graphs}.
GRASP's ID Alignment collapses record identifiers, and Decoy adds
misleading endpoint fields to disrupt structured
extraction~\cite{song2026graphs}. These defenses aim to reduce extraction but provide no formal
privacy guarantees. In contrast, our framework provides
relation-DP guarantees through document sanitization before indexing.

\noindent\textbf{Metric-DP on text.}
Prior work applies metric-DP to word
embeddings~\cite{feyisetan2020privacy}, sentence
representations~\cite{bollegala2025metric}, and LLM-generated
substitutes~\cite{awon2025clusant}.
Pr$\epsilon\epsilon$mpt combines metric-DP with format-preserving
encryption for sensitive prompt values~\cite{chowdhury2025pr},
while TDP substitutes sensitive entities in LLM
contexts~\cite{yu2026protecting}. These methods protect textual units rather than explicitly
addressing both the semantics and existence of relationships
between entities. Our framework targets relation-level privacy,
combining semantic replacement and relation deletion to protect
relation meaning and mitigate connectivity leakage.

\section{Problem Setup}
\label{sec:problem}

\subsection{Semantic Distance between Relations}
\label{sec:relation_distance}

We define an embedding-based semantic distance between relation
predicates for the privacy guarantee in Section~\ref{sec:metric_srdp}.
We also describe the extension used to represent connectivity
perturbation.

\smallskip
\noindent\textbf{Distance between predicates.}
Let $\mathcal R$ denote the domain of relations, and let
$\operatorname{Emb}:\mathcal R\to\mathbb R^p$ be a fixed encoder
that maps each relation to a finite, nonzero embedding vector
\cite{ushio2021distilling,zargayouna2017exploring}. For each
$r\in\mathcal R$, we define its normalized embedding as
\begin{equation}
z(r)=\frac{\operatorname{Emb}(r)}
{\|\operatorname{Emb}(r)\|_2}.
\label{eq:normalized_relation_embedding}
\end{equation}
The semantic distance between two predicates $r,r'\in\mathcal R$
is defined as
\begin{equation}
d_{\mathcal R}(r,r')
=\frac{1}{2}\|z(r)-z(r')\|_2.
\label{eq:relation_distance}
\end{equation}
Since the embeddings are unit-normalized, this distance depends
only on their directions and is independent of their magnitudes.
Furthermore, $d_{\mathcal R}(r,r')\in[0,1]$, with smaller values
indicating greater embedding-based semantic similarity and larger
values indicating greater semantic dissimilarity. The scaled Euclidean
distance satisfies the triangle inequality, as shown in
Appendix~\ref{appendix-distance-proof}. It is a metric when
$r\mapsto z(r)$ is injective and otherwise a pseudometric; the
privacy proof covers both cases.

\smallskip
\noindent\textbf{Extended distance between triples under connectivity perturbation.}
In addition to perturbing relation semantics, our defense may remove
edges from the relation graph to protect connectivity privacy. To capture such structural perturbation, we introduce a special
marker $\texttt{null}$ to represent the absence of an edge between
the subject $s$ and object $o$. Specifically, for a triple
$t=(s,r,o)$ and its edge-removed counterpart
$t'=(s,\texttt{null},o)$, we define their distance as
\begin{equation}
d_{\mathcal T}(t,t')
=\lambda,
\qquad \lambda>0,
\label{eq:relation_removal_distance}
\end{equation}
where $\lambda$ is a predefined positive constant that quantifies
the distance associated with connectivity perturbation. We assign a constant distance $\lambda$  because it separates the cost of structural changes from semantic differences and provides a uniform measure of the
privacy cost associated with removing any edge. The value of $\lambda$
controls the relative distance assigned to edge removal compared with
relation substitution and is fixed across all relations to ensure a
consistent distance definition.

\subsection{Relation Differential Privacy}
\label{sec:metric_srdp}

While conventional DP \cite{dwork2006calibrating,dwork2014algorithmic} provides a rigorous framework for limiting the information revealed about private data, its standard formulation typically assigns the same privacy bound to all neighboring inputs. This uniform treatment does not distinguish between relation substitutions that preserve similar semantics and those that substantially change the meaning of a knowledge graph fact.
On the other hand, metric-DP generalizes conventional DP by incorporating a distance function over inputs, allowing the privacy guarantee to vary with the distance between them. Motivated by this flexibility, we introduce {\em relation differential privacy (relation-DP)}, which instantiates the metric-DP principle for relations by defining privacy in terms of semantic distances between relations.

Let \(\mathcal R\) denote the domain of relation predicates, and let \(\mathcal O=\mathcal R\cup\{\texttt{null}\}\) be the output space, where \(\texttt{null}\) is a special marker indicating the absence of an edge. We define \emph{\(\epsilon\)-Relation Differential Privacy} as follows.

\begin{definition}[\(\epsilon\)-Relation Differential Privacy]
\label{def:metric_srdp}
A randomized mechanism \(\mathcal M:\mathcal R\to\mathcal O\) satisfies \(\epsilon\)-relation-DP if, for every pair of input relations \(r,r'\in\mathcal R\) and every output event \(\mathcal S\subseteq\mathcal O\),
\begin{equation}
\Pr[\mathcal M(r)\in\mathcal S]
\leq
e^{\epsilon d_{\mathcal R}(r,r')}
\Pr[\mathcal M(r')\in\mathcal S],
\label{eq:relation_dp}
\end{equation}
where \(d_{\mathcal R}\) is the relation distance defined in Eq.~\eqref{eq:relation_distance}.
\end{definition}

Relation-DP bounds the distinguishability of both released predicates
and retention or deletion outcomes under alternative predicates on
a fixed entity pair. The guarantee provides tighter indistinguishability bounds for semantically closer input relations and correspondingly weaker bounds for more distant relations.

\smallskip
\noindent{\bf Document-level extension.}
We extend \(\epsilon\)-relation-DP from individual relations to the collection of relations extracted from a document. Let \(D\) and \(D'\) be two private documents, and let
\(\mathcal F(D)=(t_i)_{i=1}^{m}\) and
\(\mathcal F(D')=(t_i')_{i=1}^{m}\) denote their ordered canonical triples, where \(t_i=(s_i,r_i,o_i)\) and \(t_i'=(s_i,r_i',o_i)\) for all $i\in\{1,\ldots,m\}$. We compare documents that differ only in the protected relation expressions, with all remaining text and occurrence mappings unchanged.
We define the distance between the two collections of triples as the sum of the distances between their corresponding relations:
\begin{equation}
d_{\mathcal F}(\mathcal F(D),\mathcal F(D'))
=\sum_{i=1}^{m}d_{\mathcal R}(r_i,r_i').
\label{eq:document_relation_distance}
\end{equation}

Suppose a randomized mechanism independently perturbs each relation using an \(\epsilon\)-relation-DP mechanism and reconstructs the perturbed triples while preserving their subject and object entities. Then, for every such pair \(D,D'\) and every joint output event \(\mathcal S\), the resulting mechanism \(\mathcal M\) satisfies
\begin{equation}
\Pr[\mathcal M(D)\in\mathcal S]
\leq
e^{\epsilon d_{\mathcal F}(\mathcal F(D),\mathcal F(D'))}
\Pr[\mathcal M(D')\in\mathcal S].
\label{eq:document_relation_dp}
\end{equation}
Thus, the document-level privacy guarantee follows from the individual relation-DP guarantees through composition, with the overall privacy bound determined by the sum of the distances between corresponding relations. The composition proof is provided in Appendix~\ref{appendix-reconstruction-proof}.

\subsection{Threat Model and Defense Goal}
\label{sec:threat_model}

\smallskip
{\bf Attack goal.}
Given a private document $D$ and a target entity $\ts$
appearing in it, the adversary
aims to infer sensitive information associated with $\ts$ from GraphRAG
answers. Specifically, the attack targets two types of information:
\begin{itemize}
    \item \emph{Connectivity leakage}: The adversary aims to identify
    which object entities in $D$ are directly connected to $\ts$.

    \item \emph{Relation semantics leakage}: The adversary aims to infer
    the semantic meaning of the relationship between $\ts$ and each
    directly connected object entity. An inference is considered
    successful if the inferred relation is semantically equivalent to
    the true relation, even if its exact representation differs.
\end{itemize}

\smallskip {\bf  Adversary capabilities.}
We consider a strong, knowledge-rich adversary who knows the identities
of all object entities appearing in the private document $D$, but has no
prior knowledge of the relationships among them. This assumption
reflects scenarios in which entity identities are publicly available
or already known to the adversary. It also establishes a conservative
threat model in which privacy protection does not rely on concealing
the entity set, but instead focuses on protecting sensitive relationships
among known entities.

Although the adversary is strong in terms of prior knowledge, we assume
that it interacts with GraphRAG as an external user and therefore has
only black-box access to the service. This setting reflects practical
deployments in which the underlying document, graph index, retrieved
context, and model parameters are not exposed to users. The adversary
can submit natural-language queries, observe generated answers, and
adapt subsequent queries based on information accumulated from previous
responses. However, it cannot directly inspect the private document or
the system's internal components, nor can it modify the document or
index. Document poisoning and system compromise are outside our threat
model.

\begin{figure}[t]
    \centering
    \includegraphics[width=\linewidth]{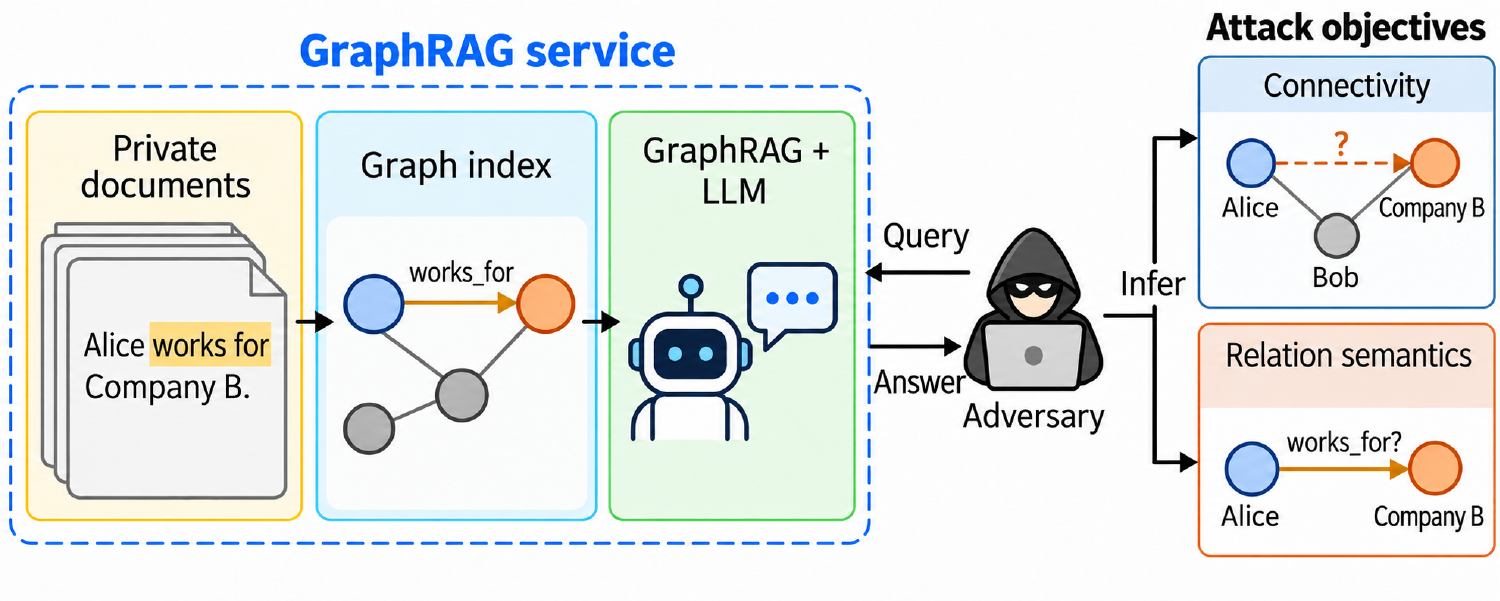}
    \vspace{-0.3cm}
    \caption{Threat model. }
    \label{fig:threat_model}
\end{figure}

Figure~\ref{fig:threat_model} illustrates the threat model. Suppose a
private document states that Alice works for Company B, where Alice is
the target entity $\ts$. The adversary knows the identities of Alice
and Company B, as well as all other object entities appearing in the
private document, but has no prior knowledge of the relationships among
them. Through black-box queries to the GraphRAG system, the adversary
may infer whether Alice is connected to Company B and, if so, the
semantic meaning of their relationship. For example, an answer
describing Alice as an employee of Company B reveals both the
{\tt Alice--Company} B connection and its employment semantics, even without
explicitly reproducing the predicate \texttt{works\_for}. By issuing
adaptive follow-up queries, the adversary may attempt to identify all
object entities directly connected to Alice and infer the semantic
meaning of each corresponding relationship, thereby reconstructing
Alice's entire one-hop neighborhood.

\smallskip
\noindent{\bf Defense objective.}
We aim to design a defense mechanism that mitigates the disclosure of sensitive relations involving $\ts$ along two complementary dimensions: {\bf structural privacy}, which limits the ability to infer whether $\ts$ is connected to other entities, and {\bf semantic privacy}, which limits the ability to infer the meanings of these relations. At the same time, the defense should preserve the utility of the protected document for downstream question answering (QA). To achieve these objectives, we develop a relation-level perturbation mechanism guided by the relation-DP guarantee (Definition~\ref{def:metric_srdp}), with the goal of protecting sensitive relational information while retaining sufficient semantic and structural information for accurate QA.

\section{Methodology}
\label{sec:method}

\begin{figure*}[t]
    \centering
    \includegraphics[width=0.84\textwidth]{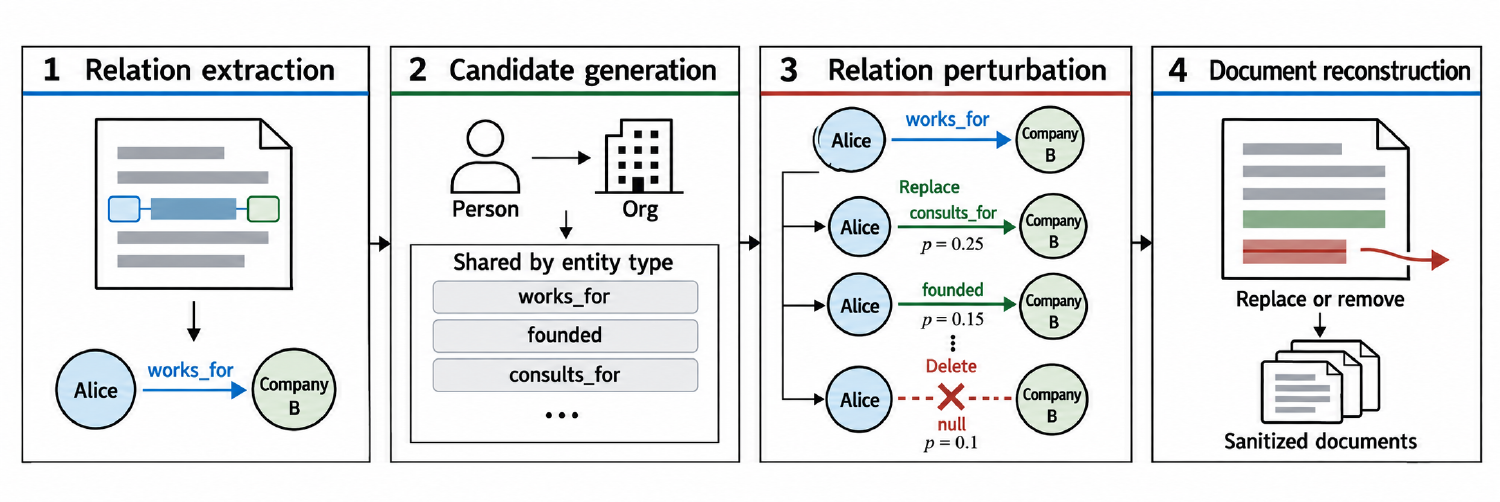}
    \vspace{-0.3cm}
    \caption{Overview of \system.}
    \label{fig:overview}
\end{figure*}

Given a private document $D$ and privacy budget $\epsilon$, \system\ generates a sanitized document $\widetilde D$ for GraphRAG indexing and query answering. Figure~\ref{fig:overview} shows four steps: (1) extract canonical entity--relation triples and record their occurrences; (2) generate type-conditioned candidate relations; (3) perturb relations using these candidates and \texttt{null}; and (4) reconstruct $\widetilde D$ from the perturbed triples. We describe each step below.

\subsection{Relation Extraction}
\label{sec:relation_extraction}

The first step of \system\ is to extract subject--relation--object triples from the private document $D$. We use a large language model (LLM) for this task \cite{huguet2021rebel,mo2025kggen}. The LLM extracts only relations supported by the document, excluding unsupported, negated, or hypothetical relations. The extraction prompt is provided in Appendix~\ref{appendix-prompt}.

To ensure consistent representation, entity aliases and unambiguous coreferential mentions are mapped to the same entity identity, and semantically equivalent relation expressions are normalized into \texttt{snake\_case} predicates while preserving subject--object direction. For example, both \emph{Alice works at Company B} and \emph{Alice is employed by Company B} are normalized to the same triple, $(`\text{Alice}',\texttt{works\_for},`\text{Company B}')$.

\subsection{Relation Candidate Generation}
\label{sec:candidate_generation}

Next, \system\ generates candidate relations for perturbing each extracted triple. The key idea is to construct candidate pools based on the types of the subject and object entities, rather than the original relation. Specifically, \system\ follows two steps to generate the candidates: (1) {\bf Entity typing}: it assigns each entity a type from a fixed inventory; (2) {\bf Candidate generation}: it generates a pool of candidate predicates for each observed relation based on the types of its subject and object entities. All triples with the same subject and object entities use the same candidate pool. This design makes generated relation candidates independent of the extracted ones  and avoids generating a separate pool for every triple. Next, we explain the details of the two steps.

\smallskip
{\bf Entity typing.}
Given the set of canonical entities $V$ appearing in the extracted triples, we define a type assignment function $\tau:V\to\mathcal U$ that assigns each entity a type from a fixed inventory. Specifically, we adopt an entity-type inventory based on OntoNotes~\cite{weischedel2013ontonotes}:
\begin{equation}
\begin{aligned}
\mathcal U=\{&
\textsc{Person},\textsc{Norp},\textsc{Fac},\textsc{Org},
\textsc{Gpe},\textsc{Loc},\textsc{Product},\textsc{Event},\\
&\textsc{Work\_Of\_Art},\textsc{Law},\textsc{Language}\}.
\end{aligned}
\end{equation}
Each entity is assigned a type based on its canonical identity and the fixed inventory, without using source sentences or extracted relation predicates. Given a subject entity $s$ and an object entity $o$, we define their {\em ordered type pair} as
\[
g(s,o)=(\tau(s),\tau(o)),
\]
where the subject type always comes first and the object type second. This ordering is essential because relation predicates are directional: reversing the subject and object types can change the set of type-compatible predicates. For example, \texttt{works\_for} describes a relation from a person to an organization, whereas \texttt{employs} describes the corresponding relation in the reverse direction. Accordingly, these two directions are treated as distinct type pairs for candidate generation. In our running example, \texttt{Alice} is assigned type \textsc{Person} and \texttt{Company B} type \textsc{Org}, yielding $g(\texttt{Alice},\texttt{Company B})=(\textsc{Person},\textsc{Org})$.

\smallskip
\textbf{Candidate generation.}
For each extracted triple $(s,r,o)$, we prompt an LLM to generate a pool of $K$ distinct candidate relations for $r$, denoted by $\mathcal{C}_{r} = \{r_1, \ldots, r_K\}$, where $K \geq 1$ is a user-specified parameter. The candidate relations must satisfy the following three requirements.
\begin{enumerate}[leftmargin=*,itemsep=2pt]
    \item \textbf{Type compatibility}: Each $r_i \in \mathcal{C}_{r}$ must represent a plausible relation between entity types $\tau(s)$ and $\tau(o)$. For example, for the type pair $(\textsc{Person}, \textsc{Org})$, plausible relations include \texttt{works\_for}, \texttt{founded}, and \texttt{invests\_in}, whereas \texttt{headquartered\_in} is incompatible.

    \item \textbf{Semantic distinctness}: The candidate relations must be semantically distinct. Relations that are semantically equivalent, such as synonyms or paraphrases, are merged into a single canonical relation. For example, \texttt{works\_for} and \texttt{employed\_by} are treated as equivalent, whereas \texttt{works\_for} and \texttt{founded} represent distinct relations.

    \item \textbf{Non-subsumption}: No candidate relation may subsume another relation in the same pool. If one relation logically implies another, the broader relation is removed in favor of the more specific one. For example, \texttt{spouse\_of} implies \texttt{family\_member\_of}; therefore, the broader \texttt{family\_member\_of} relation is removed, while the more specific \texttt{spouse\_of} relation is retained.
\end{enumerate}
The prompt used for generating candidate relations is provided in Appendix~\ref{appendix-prompt}. Given the generated candidate relations, for each extracted triple $t=(s,r,o)$, we construct its candidate pool (denoted as $\mathcal{C}_t$) as
\[
\mathcal{C}_t = \{(s,r_i,o) \mid r_i \in C_{r}\}.\]
Thus, all triples sharing the same subject and object entities use the same candidate pool.

\subsection{Relation Perturbation}
\label{sec:relation_perturbation}

At this stage, \system\ protects both relation semantics and connectivity by randomly replacing each relation with a candidate from the pool or deleting it.
An exponential mechanism governs these choices by balancing
semantic preservation against privacy protection through two
parameters: $\epsilon$, which controls the randomness of the
perturbation, and $\lambda$, which controls the propensity for
deletion.
Algorithm~\ref{alg:relation_mem} summarizes the relation perturbation
procedure. We explain the details below.

For each triple $t(s,r,o)$ retrieved from the private document $D$, let $\mathcal{C}_r$ be its candidate pool (Section~\ref{sec:candidate_generation}).
To perturb both relation meaning and released connectivity, we
augment this pool with \texttt{null}:
$\mathcal{O}_r=\mathcal C_r\cup\{\texttt{null}\}$.
Selecting a real candidate replaces
the original relation, whereas selecting \texttt{null} removes the
relation from the released representation.

We assign each candidate outcome a utility score based on its semantic distance from the original relation. Specifically, for an original relation $r$ and a candidate outcome $r'\in\mathcal{C}_r$, we define the utility function $u(r,r')$ for replacing $r$ with $r'$ as follows:
\begin{equation}
u(r, r')=
\begin{cases}
-d_{\mathcal R}(r,r'), & r'\in\mathcal C_r,\\
-\lambda,
& r'=\texttt{null}.
\end{cases}
\label{eq:relation_release_utility}
\end{equation}
where
$d_{\mathcal R}$ is defined in Eq.~\eqref{eq:relation_distance} and $\lambda$ is defined in
Eq.~\eqref{eq:relation_removal_distance}.

For a real candidate, the utility is the negative semantic distance,
so candidates that preserve more of the original relation's meaning
receive higher scores. In contrast, \texttt{null} represents deletion
and has no predicate semantics; we therefore assign it the constant
utility $-\lambda$, independent of the input relation.

The removal distance $\lambda$ is fixed across compared inputs but
remains a configurable parameter. For a fixed $\epsilon>0$ and
candidate pool, decreasing $\lambda$ increases the probability of
selecting \texttt{null}, allowing more relations to be removed. This
parameter therefore controls the trade-off between privacy protection
and question-answering (QA) utility, which we investigate in
Section~\ref{sec:rq3}.

For each relation $r$, we independently sample a perturbed
outcome $\widetilde r\in\mathcal O_i$ using the exponential
mechanism:
\begin{equation}
\Pr[\widetilde r=r'\mid r]
=
\frac{\exp(\epsilon u(r,r')/2)}
{\sum_{z\in\mathcal O_i}\exp(\epsilon u(r,z)/2)}.
\label{eq:relation_mem}
\end{equation}

The denominator normalizes the scores over all possible outcomes,
including both relation replacements and deletion. Consequently,
both types of perturbation are governed by the same probability
distribution, with their relative probabilities determined by their
respective utility scores. Furthermore, a smaller $\epsilon$
produces a flatter sampling distribution, increasing randomness,
whereas a larger $\epsilon$ places more probability on outcomes with
higher utility.

\begin{algorithm}[t]
\caption{Relation Perturbation}
\label{alg:relation_mem}
\KwIn{Ordered triples $\mathcal F(D)$; fixed pools
$\{\mathcal C_i\}$ with $|\mathcal C_i|=K\geq1$; distance $d_{\mathcal R}$;
$\lambda\geq0$; $\epsilon\geq0$}
\KwOut{Perturbed triples $\widetilde{\mathcal F}(D)$ in input order}
$\widetilde{\mathcal F}(D)\leftarrow[\,]$\;
\For{$i=1,\ldots,m$}{
    $\mathcal O_i\leftarrow\mathcal C_i\cup\{\texttt{null}\}$\;
    \ForEach{$c\in\mathcal C_i$}{
        $w_i(c)\leftarrow\exp(-\epsilon d_{\mathcal R}(r_i,c)/2)$\;
    }
    $w_i(\texttt{null})\leftarrow\exp(-\epsilon\lambda/2)$\;
    Sample $\widetilde r_i\in\mathcal O_i$ with probability
    $w_i(\widetilde r_i)/\sum_{y\in\mathcal O_i}w_i(y)$\;
    Append $(s_i,\widetilde r_i,o_i)$ to
    $\widetilde{\mathcal F}(D)$\;
}
\Return{$\widetilde{\mathcal F}(D)$}\;
\end{algorithm}

We have the following theorem:

\begin{theorem}
\label{thm:relation_mem_privacy}
Algorithm~\ref{alg:relation_mem} satisfies
$\epsilon$-relation-DP (Definition~\ref{def:metric_srdp}).
\end{theorem}

The proof of Theorem~\ref{thm:relation_mem_privacy} can be found in
Appendix~\ref{appendix-distance-proof}.

\subsection{Document Reconstruction}
\label{sec:reconstruction}
Next, we incorporate the perturbed relations into the source document
before indexing. Since a subject--relation--object triple may appear
multiple times with different wording, we record all occurrences of
each triple $t_i$ in the input document to ensure that its perturbed
relation is applied consistently across all mentions:
\[
\mathcal P_i=\{(p_{ij},\ell_{ij})\}_{j=1}^{k_i},
\]
where $k_i$ is the number of occurrences of $t_i$, $p_{ij}$ identifies the
sentence containing occurrence $j$, and $\ell_{ij}$ records the exact
text span expressing the relation. These records are collected during
extraction (Section~\ref{sec:relation_extraction}). Each canonical
triple receives one sampled outcome, shared across all its occurrences.

To reconstruct the sanitized document, we first replace each recorded
relation expression in $D$ with an indexed placeholder, producing a
masked document $A(D)$. For example, \emph{Alice works at Company B}
becomes \emph{Alice \texttt{[REL\_i]} Company B}, where
\texttt{[REL\_i]} corresponds to triple $i$. The occurrence records
$\mathcal P_i$ locate these placeholders and their sentences for
subsequent replacement or deletion. All other text remains unchanged.
The compared documents in Section~\ref{sec:metric_srdp} differ only in
the protected relation expressions and therefore yield the same masked
document, denoted by $A$.

If a triple receives \texttt{null}, we delete every sentence containing
a recorded occurrence of that triple. A sentence containing multiple
triples is deleted if any of them receives \texttt{null}.
In the remaining sentences, we fill each placeholder with a
verbalization of its sampled relation $\widetilde r_i$, using the masked
document without consulting the original relation expression. For
example, replacing \texttt{works\_for} with \texttt{founded} changes
\emph{Alice works at Company B} to \emph{Alice founded Company B}.
All surviving mentions of the same triple use the same sampled
relation. Other text and sentence order remain unchanged. The resulting
sanitized document $\widetilde D$ is indexed once for subsequent queries
without modifying GraphRAG's retrieval or generation procedures.

\begin{theorem}
\label{thm:reconstruction_privacy}
The reconstructed document satisfies the document-level
relation-DP guarantee (Eq.~\eqref{eq:document_relation_dp}).
\end{theorem}

The result follows the composition and
postprocessing property of DP~\cite{dwork2014algorithmic}; the detailed proof is provided in
Appendix~\ref{appendix-reconstruction-proof}.

\section{Experiments}
\label{sec:experiments}
\begin{table*}[!t]
\centering
\caption{Performance of \system\
 and defense baselines ($\epsilon=5$, GPT-4o-mini). We use GRASP and targeted
GraphSteal to measure the privacy leakage of connectivity and relation.
Bold values mark the lowest recovery F1 for each system, dataset,
attack, and metric, and the highest QA F1 for each system and dataset.}
\label{tab:defense_comparison}
\small
\setlength{\tabcolsep}{3pt}
\renewcommand{\arraystretch}{0.85}
\resizebox{\textwidth}{!}{
\begin{tabular}{ll*{10}{r}}
\toprule
\multirow{3}{*}{\textbf{Victim}} &
\multirow{3}{*}{\textbf{Defense}} &
\multicolumn{5}{c}{\textbf{2WikiMultiHopQA}} &
\multicolumn{5}{c}{\textbf{HotpotQA}} \\
\cmidrule(lr){3-7}\cmidrule(lr){8-12}
& & \multicolumn{2}{c}{GRASP} & \multicolumn{2}{c}{GraphSteal}
& \multirow{2}{*}{QA F1 $\uparrow$}
& \multicolumn{2}{c}{GRASP} & \multicolumn{2}{c}{GraphSteal}
& \multirow{2}{*}{QA F1 $\uparrow$} \\
\cmidrule(lr){3-4}\cmidrule(lr){5-6}\cmidrule(lr){8-9}\cmidrule(lr){10-11}
& & Conn. F1 $\downarrow$ & Rel. F1 $\downarrow$
& Conn. F1 $\downarrow$ & Rel. F1 $\downarrow$ &
& Conn. F1 $\downarrow$ & Rel. F1 $\downarrow$
& Conn. F1 $\downarrow$ & Rel. F1 $\downarrow$ & \\
\midrule
\multirow{6}{*}{LightRAG} & no-defense & 89.63 & 84.26 & 89.14 & 85.54 & \textbf{49.48} & 83.78 & 75.97 & 90.31 & 80.98 & 61.47 \\
 & Abstractive & 88.76 & 83.33 & 89.72 & 85.13 & 46.37 & 82.81 & 75.80 & 89.51 & 80.65 & 60.41 \\
 & ID Alignment & 87.28 & 83.36 & 89.33 & 85.88 & 46.53 & 83.40 & 75.32 & 90.70 & 80.04 & 60.24 \\
 & Decoy & 85.57 & 80.06 & \textbf{82.57} & 77.87 & 49.09 & 80.56 & 74.07 & 78.99 & 73.26 & 61.23 \\
 & Protective Prompt & 86.93 & 83.61 & 88.89 & 84.69 & 48.17 & 82.73 & 75.60 & 89.09 & 80.82 & \textbf{61.86} \\
 & \system\ (ours) & \textbf{83.30} & \textbf{40.94} & 84.47 & \textbf{44.31} & 44.85 & \textbf{71.24} & \textbf{37.64} & \textbf{73.51} & \textbf{34.86} & 58.49 \\
\midrule
\multirow{6}{*}{PathRAG} & no-defense & 79.31 & 72.71 & 82.49 & 77.20 & \textbf{55.39} & 74.38 & 67.84 & 82.66 & 72.02 & \textbf{64.47} \\
 & Abstractive & 79.25 & 72.37 & 81.85 & 76.79 & 51.30 & 72.57 & 63.19 & 82.13 & 71.43 & 61.65 \\
 & ID Alignment & 76.58 & 71.05 & 79.84 & 74.67 & 52.06 & 73.92 & 63.42 & 82.41 & 72.07 & 63.20 \\
 & Decoy & 78.33 & 72.23 & 79.11 & 69.70 & 52.99 & 73.71 & 67.19 & 78.79 & 68.09 & 61.91 \\
 & Protective Prompt & 79.27 & 71.42 & 81.81 & 77.00 & 50.47 & 73.10 & 66.38 & 82.10 & 70.64 & 64.45 \\
 & \system\ (ours) & \textbf{75.93} & \textbf{37.93} & \textbf{78.86} & \textbf{42.42} & 52.40 & \textbf{68.51} & \textbf{36.21} & \textbf{74.04} & \textbf{40.65} & 60.85 \\
\midrule
\multirow{6}{*}{HippoRAG 2} & no-defense & 76.99 & 68.51 & 75.71 & 69.93 & \textbf{70.50} & 70.92 & 62.48 & 78.37 & 67.21 & 71.91 \\
 & Abstractive & 73.34 & 65.05 & 69.94 & 53.30 & 70.27 & 70.62 & 62.60 & 68.71 & 54.06 & 71.98 \\
 & ID Alignment & 71.27 & 67.61 & 75.34 & 66.49 & 69.33 & 70.68 & 62.13 & 74.95 & 65.51 & \textbf{72.03} \\
 & Decoy & 70.19 & 63.96 & 74.92 & 67.99 & 68.86 & 54.75 & 48.00 & 71.29 & 59.60 & 71.19 \\
 & Protective Prompt & 76.25 & 67.01 & 75.15 & 66.67 & 68.94 & 68.81 & 58.88 & 77.24 & 64.84 & 71.14 \\
 & \system\ (ours) & \textbf{69.75} & \textbf{19.61} & \textbf{69.60} & \textbf{25.81} & 61.53 & \textbf{45.03} & \textbf{13.63} & \textbf{63.42} & \textbf{23.47} & 67.25 \\
\bottomrule
\end{tabular}
}
\end{table*}

We organize our evaluation around three research questions:
\begin{ditemize}
    \item \textbf{RQ1:} How does \system\ compare with existing
    defenses in protecting GraphRAG against reconstruction attacks on
    connectivity and relation semantics?

    \item \textbf{RQ2:} What trade-off does \system\ achieve between
    privacy protection and downstream question-answering (QA) utility?

    \item \textbf{RQ3:} How do different parameter settings affect
    the privacy protection and QA utility of \system?
\end{ditemize}

\subsection{Experimental Setup}
\label{sec:exp_setup}
We run experiments on a server with an NVIDIA A100 GPU with 40 GB
of GPU memory. We implement the algorithms in Python using PyTorch
and report the mean over three trials.

\noindent\textbf{Data, models, and GraphRAG systems.}
We evaluate privacy protection and utility on
HotpotQA~\cite{yang2018hotpotqa} and
2WikiMultiHopQA~\cite{ho2020constructing}.
Their source documents form the protected corpus, while validation
questions and reference answers are used for QA evaluation. We evaluate three mainstream GraphRAG systems:
LightRAG~\cite{guo2025lightrag}, PathRAG~\cite{chen2026pathrag}, and
HippoRAG 2~\cite{gutierrez2025rag}. The victim generators are two
closed-source models, GPT-4o-mini~\cite{openai2024gpt4omini} and
Gemini-2.5-Flash. Results for Gemini-2.5-Flash appear in
Appendix~\ref{appendix-gemini-results}.
We use GPT-5-mini to perform relation extraction and candidate generation
in the DP-protected document reconstruction pipeline.
We encode relation predicates using \texttt{nli-roberta-base-v2}.
Unless otherwise specified, we use $K=10$ and $\lambda=1.0$.

\noindent\textbf{Recovery Attacks.}
We employ two state-of-the-art black-box reconstruction attacks against GraphRAG systems:  \textbf{GRASP}~\cite{song2026graphs}
and \textbf{GraphSteal}~\cite{gu2026graphsteal}. Both attacks recover both entity connections
and relation semantics within the target entity's one-hop
subgraph.
Appendix~\ref{appendix-attack-details} provides the attack settings.

\noindent\textbf{Privacy metrics.}
We measure \emph{Connection Recovery (Conn.)} and
\emph{Relation Recovery (Rel.)}, following the distinction between
endpoint and relation matching~\cite{song2026graphs}.
\emph{Conn.} checks whether the attacker
identifies the correct directed entity pair, regardless of the
predicted predicate. Entity aliases are mapped to canonical names
before comparison. \emph{Rel.} additionally
uses GPT-4o-mini to judge whether the predicted predicate captures
the true relation's meaning.
Appendix~\ref{appendix-privacy-metrics} defines the matching rules
and precision, recall, and F1 calculations.

\noindent\textbf{Utility metrics.}
Following prior graph-based RAG evaluations
\cite{gutierrez2024hipporag,gutierrez2025rag}, we measure QA utility
using token-level F1, the harmonic mean of token precision and
recall. We follow each benchmark's answer and
alias handling, average scores over questions, and report percentages.
Higher scores indicate better utility.

\noindent\textbf{Defense baselines.}
We compare \system\ with four existing GraphRAG defenses: {\em Abstractive}
\cite{liu2026exposing}, {\em ID Alignment}, {\em Decoy}
\cite{song2026graphs}, and {\em Protective Prompt}
\cite{gu2026graphsteal}. Abstractive restricts responses to
paraphrased query-relevant evidence. ID Alignment collapses retrieved
record identifiers, while Decoy introduces misleading endpoint fields.
Protective Prompt adds a
system-level instruction against disclosing retrieved knowledge-base
content. Appendix~\ref{appendix-defense-baselines} provides details of these methods.

\begin{table*}[t]
\centering
\small
\caption{Relation privacy leakage and QA utility of \system\ under
different privacy budgets (GPT-4o-mini).}
\label{tab:privacy_utility_budgets}
\setlength{\tabcolsep}{4pt}
\renewcommand{\arraystretch}{0.9}
\resizebox{\textwidth}{!}{
\begin{tabular}{ll*{10}{r}}
\toprule
\multirow{3}{*}{\textbf{Victim}} &
\multirow{3}{*}{\textbf{Config}} &
\multicolumn{5}{c}{\textbf{2WikiMultiHopQA}} &
\multicolumn{5}{c}{\textbf{HotpotQA}} \\
\cmidrule(lr){3-7}\cmidrule(lr){8-12}
& & \multicolumn{2}{c}{GRASP} & \multicolumn{2}{c}{GraphSteal}
& \multirow{2}{*}{QA F1 $\uparrow$}
& \multicolumn{2}{c}{GRASP} & \multicolumn{2}{c}{GraphSteal}
& \multirow{2}{*}{QA F1 $\uparrow$} \\
\cmidrule(lr){3-4}\cmidrule(lr){5-6}\cmidrule(lr){8-9}\cmidrule(lr){10-11}
& & Conn. F1 $\downarrow$ & Rel. F1 $\downarrow$
& Conn. F1 $\downarrow$ & Rel. F1 $\downarrow$ &
& Conn. F1 $\downarrow$ & Rel. F1 $\downarrow$
& Conn. F1 $\downarrow$ & Rel. F1 $\downarrow$ & \\
\midrule
\multirow{5}{*}{HippoRAG 2} & no-DP & 76.99 & 68.51 & 75.71 & 69.93 & 70.50 & 70.92 & 62.48 & 78.37 & 67.21 & 71.91 \\
 & $\epsilon=10$ & 76.39 & 20.14 & 74.47 & 29.70 & 62.99 & 55.35 & 19.31 & 71.77 & 30.99 & 67.45 \\
 & $\epsilon=5$ & 69.75 & 19.61 & 69.60 & 25.81 & 61.53 & 45.03 & 13.63 & 63.42 & 23.47 & 67.25 \\
 & $\epsilon=1$ & 59.39 & 18.12 & 54.08 & 21.79 & 53.40 & 44.82 & 11.37 & 62.67 & 21.71 & 64.90 \\
 & $\epsilon=0.1$ & 49.73 & 11.01 & 52.22 & 14.87 & 52.75 & 43.92 & 10.09 & 59.76 & 20.58 & 63.30 \\
\midrule
\multirow{5}{*}{LightRAG} & no-DP & 89.63 & 84.26 & 89.14 & 85.54 & 49.48 & 83.78 & 75.97 & 90.31 & 80.98 & 61.47 \\
 & $\epsilon=10$ & 88.79 & 47.26 & 89.09 & 46.32 & 46.73 & 83.07 & 50.60 & 87.97 & 47.63 & 58.56 \\
 & $\epsilon=5$ & 83.30 & 40.94 & 84.47 & 44.31 & 44.85 & 71.24 & 37.64 & 73.51 & 34.86 & 58.49 \\
 & $\epsilon=1$ & 65.45 & 28.71 & 65.57 & 31.42 & 45.17 & 68.09 & 35.59 & 70.96 & 34.27 & 57.24 \\
 & $\epsilon=0.1$ & 59.01 & 25.41 & 63.78 & 27.59 & 45.06 & 66.23 & 35.01 & 68.65 & 33.78 & 57.23 \\
\midrule
\multirow{5}{*}{PathRAG} & no-DP & 79.31 & 72.71 & 82.49 & 77.20 & 55.39 & 74.38 & 67.84 & 82.66 & 72.02 & 64.47 \\
 & $\epsilon=10$ & 78.77 & 42.65 & 81.48 & 43.84 & 54.24 & 71.45 & 40.25 & 81.05 & 45.49 & 62.58 \\
 & $\epsilon=5$ & 75.93 & 37.93 & 78.86 & 42.42 & 52.40 & 68.51 & 36.21 & 74.04 & 40.65 & 60.85 \\
 & $\epsilon=1$ & 56.59 & 23.16 & 64.84 & 32.50 & 47.54 & 62.69 & 35.88 & 69.17 & 40.44 & 60.49 \\
 & $\epsilon=0.1$ & 58.29 & 21.30 & 62.68 & 30.92 & 46.90 & 57.97 & 30.97 & 66.26 & 39.92 & 58.95 \\
\bottomrule
\end{tabular}
}
\end{table*}

\subsection{RQ1: Comparison with Existing Defenses}
\label{sec:rq1}

Table~\ref{tab:defense_comparison} compares \system\ at $\epsilon=5$
with four existing defenses and no-defense on two datasets across
three GraphRAG systems, all using GPT-4o-mini as the generator.
We measure privacy leakage through connection and relation recovery
F1 under GRASP and targeted GraphSteal, and evaluate QA utility using
QA F1.

\noindent\textbf{Relation recovery.}
\system\ achieves the lowest Rel. F1 and the largest reduction from
no-defense in all 12 system--dataset--attack combinations.
Its Rel. F1 is 26.98--44.35 percentage points below even the strongest
baseline in each setting, compared with reductions of 31.37--48.90
points from no-defense. For LightRAG on 2WikiMultiHopQA, \system\
reduces GRASP Rel. F1 to 40.94, whereas the four baselines yield
80.06--83.61. Under GraphSteal, the corresponding scores are 44.31
and 77.87--85.88.

\noindent\textbf{Connection recovery.}
\system\ also achieves the lowest Conn. F1 and the largest reduction
from no-defense in 11 of the 12 combinations, outperforming the
strongest baseline by 0.25--9.72 percentage points in these settings.
The sole exception is LightRAG on 2WikiMultiHopQA under GraphSteal,
where Decoy yields 82.57 Conn. F1 versus 84.47 for \system.
Even there, \system\ provides substantially lower Rel. F1
(44.31 versus 77.87).

\noindent\textbf{QA utility.}
Although \system\ generally yields lower QA F1 than existing
defenses, the additional utility loss is modest in most settings.
In five of the six system--dataset combinations, its gap from the
baseline with the highest QA F1 is 0.59--4.78 percentage points.
For PathRAG on 2WikiMultiHopQA, \system\ retains 52.40 F1, only
0.59 points below the best baseline's 52.99. For LightRAG on
HotpotQA, its QA F1 is 58.49, a gap of 3.37 points.
The larger gap of 8.74 points occurs for HippoRAG 2 on
2WikiMultiHopQA. Section~\ref{sec:rq2} varies the privacy budget to examine
this trade-off beyond the fixed $\epsilon=5$ comparison.

\subsection{RQ2: Privacy--Utility Trade-off}
\label{sec:rq2}

Table~\ref{tab:privacy_utility_budgets} examines the privacy--utility
trade-off for $\epsilon\in\{10,5,1,0.1\}$ across both datasets and
all three GraphRAG systems, using GPT-4o-mini as the victim generator.
No-DP uses the original document; the other rows use documents
sanitized by \system\ at the indicated budget. All scores are
percentages. Overall, smaller $\epsilon$ reduces connection and
relation recovery, strengthening privacy protection at the cost of
QA utility; larger $\epsilon$ retains more utility but permits
more recovery.

For LightRAG on 2WikiMultiHopQA, decreasing $\epsilon$ from 10 to 1
and then to 0.1 reduces GRASP Conn. F1 from 88.79 to 65.45 and
59.01, and Rel. F1 from 47.26 to 28.71 and 25.41. GraphSteal shows
the same trend: Conn. F1 decreases from 89.09 to 65.57 and 63.78,
while Rel. F1 decreases from 46.32 to 31.42 and 27.59.
QA F1 correspondingly declines from 46.73 to 45.17 and 45.06.
The trade-off is also evident for PathRAG on HotpotQA: reducing
$\epsilon$ from 10 to 0.1 lowers GRASP Conn. F1 from 71.45 to
57.97 and Rel. F1 from 40.25 to 30.97, while QA F1 falls from
62.58 to 58.95. The utility cost varies across systems; for
HippoRAG 2 on 2WikiMultiHopQA, the same budget reduction lowers
QA F1 from 62.99 to 52.75. These results demonstrate that
adjusting $\epsilon$ provides control over the balance between
relation privacy and answer quality.

\subsection{RQ3: Sensitivity Analysis}
\label{sec:rq3}

\begin{table}[t]
\centering
\small
\setlength{\tabcolsep}{1.5pt}
\renewcommand{\arraystretch}{0.88}
\caption{Null-candidate ablation on 2WikiMultiHopQA with
LightRAG under GRASP.}
\label{tab:null_ablation}
\begin{tabular*}{\linewidth}{@{\extracolsep{\fill}}lrrrrrr@{}}
\toprule
\multirow{2}{*}{$\epsilon$} &
\multicolumn{2}{c}{Conn. F1 $\downarrow$} &
\multicolumn{2}{c}{Rel. F1 $\downarrow$} &
\multicolumn{2}{c}{QA F1 $\uparrow$} \\
\cmidrule(lr){2-3}\cmidrule(lr){4-5}\cmidrule(lr){6-7}
& w/o \texttt{null} & w/ \texttt{null}
& w/o \texttt{null} & w/ \texttt{null}
& w/o \texttt{null} & w/ \texttt{null} \\
\midrule
1 & 89.84 & 65.45 & 48.39 & 28.71 & 46.97 & 45.17 \\
5 & 86.56 & 83.30 & 45.39 & 40.94 & 47.52 & 44.85 \\
\bottomrule
\end{tabular*}
\end{table}

\noindent\textbf{Effect of \texttt{null} candidate.}
Table~\ref{tab:null_ablation} compares \system\ with a variant that
excludes \texttt{null} from the output domain and renormalizes the
replacement probabilities. We use LightRAG on 2WikiMultiHopQA under
GRASP at $\epsilon\in\{1,5\}$.
At $\epsilon=1$, including \texttt{null} reduces Conn. F1 by 24.39
percentage points and Rel. F1 by 19.68 points, with an additional
QA F1 loss of 1.80 points. At $\epsilon=5$, the corresponding
reductions are 3.26 and 4.45 points, with a QA F1 loss of 2.67 points.
Thus, enabling deletion reduces both connection and relation recovery,
with larger recovery reductions at $\epsilon=1$ but an additional
utility cost at both budgets.

\begin{figure}[t]
\centering
\definecolor{candidateQA}{HTML}{385E82}
\definecolor{candidateConn}{HTML}{A15D45}
\definecolor{candidateRel}{HTML}{587A64}
{\small
\tikz[baseline=-0.5ex]{\draw[candidateQA,line width=0.8pt] (0,0)--(0.5,0);
\fill[candidateQA] (0.25,0) circle (1.5pt);}\, QA F1
\qquad
\tikz[baseline=-0.5ex]{\draw[candidateConn,line width=0.8pt] (0,0)--(0.5,0);
\fill[candidateConn] (0.20,-0.05) rectangle (0.30,0.05);}\, Conn. F1
\qquad
\tikz[baseline=-0.5ex]{\draw[candidateRel,line width=0.8pt] (0,0)--(0.5,0);
\fill[candidateRel] (0.25,0.06)--(0.19,-0.04)--(0.31,-0.04)--cycle;}\, Rel. F1
\par}
\vspace{0.5em}
\begin{subfigure}[t]{0.23\textwidth}
    \centering
    \includegraphics[width=\linewidth]{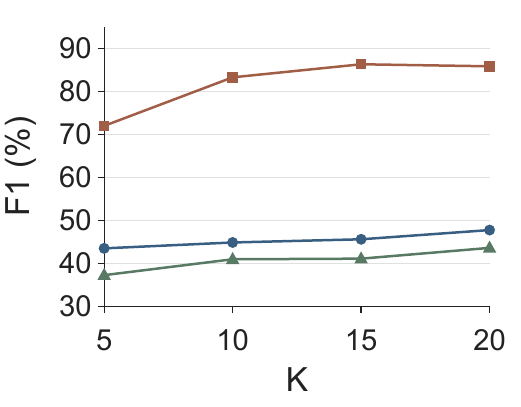}
    \caption{\# of relation candidates $K$}
    \label{fig:candidate_k_lightrag}
\end{subfigure}\hfill
\begin{subfigure}[t]{0.23\textwidth}
    \centering
    \includegraphics[width=\linewidth]{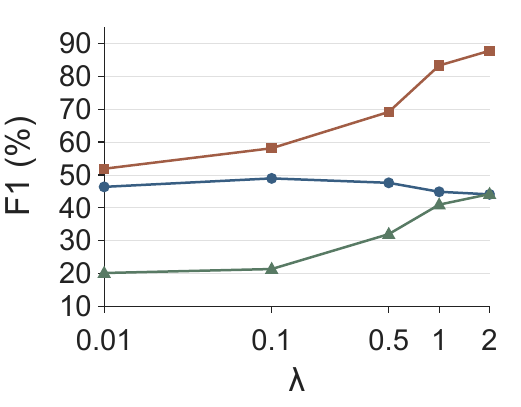}
    \caption{Removal distance $\lambda$}
    \label{fig:lambda_lightrag}
\end{subfigure}
\caption{Sensitivity analysis.}
\label{fig:sensitivity_parameters}
\end{figure}

\noindent\textbf{Number of relation candidates.}
Figure~\ref{fig:sensitivity_parameters}(a) varies
$K\in\{5,10,15,20\}$ with LightRAG on 2WikiMultiHopQA under
GRASP at $\epsilon=5$ and $\lambda=1$.
Increasing $K$ from 5 to 20 raises QA F1 from 43.48 to 47.74,
Conn. F1 from 72.00 to 85.87, and Rel. F1 from 37.24 to 43.59.
QA and relation recovery increase across the tested candidate counts, whereas
connection recovery peaks at 86.34 for $K=15$ before decreasing
slightly at $K=20$. Larger pools improve QA utility overall in
these runs but also allow greater relation recovery. Changing $K$
affects both the available replacements and null's probability
through the normalizer in Eq.~\eqref{eq:relation_mem}.

\noindent\textbf{Edge-removal distance $\lambda$.}
At fixed $\epsilon>0$ and candidate pool, a smaller $\lambda$
increases the null weight $\exp(-\epsilon\lambda/2)$ and hence
the probability of relation deletion.
Figure~\ref{fig:sensitivity_parameters}(b) varies
$\lambda\in\{0.01,0.1,0.5,1,2\}$ in the same LightRAG setting,
holding $\epsilon=5$ and the default $K=10$ candidate pool fixed.
The horizontal axis is logarithmic.
Increasing $\lambda$ from 0.01 to 2 raises Conn. F1 from 51.87 to
87.79 and Rel. F1 from 20.12 to 44.15. QA F1 first increases from
46.35 to 48.93 at $\lambda=0.1$, then decreases to 44.07 at
$\lambda=2$. Increasing $\lambda$ therefore retains more perturbed relations,
raising attack recovery across the tested values, while QA utility
peaks at $\lambda=0.1$. This pattern suggests that deleting a relation
can be preferable to retaining an incorrect replacement, which may
mislead answer generation. However, an excessively small $\lambda$
may remove too many relations and reduce the evidence available
for QA, also harming utility. At $K=10$ and $\lambda=1$,
both sweeps share QA, Conn., and Rel. F1 scores of 44.85, 83.30,
and 40.94, respectively.

\section{Conclusion}
\label{sec:conclusion}
We proposed \system\ to mitigate connectivity and relation semantics leakage through GraphRAG answers.
\system\ replaces sensitive relations based on semantic distance or deletes them, then consistently updates documents before indexing for use with GraphRAG systems. \system\ provides relation-DP guarantees over relation values,
and mitigates connectivity leakage through deletion. Experiments on two datasets across three GraphRAG systems show that,
compared with four existing defenses, \system\ substantially reduces
relation recovery with limited QA utility loss in most settings,
while also reducing connection recovery.

\clearpage
\bibliographystyle{ACM-Reference-Format}
\bibliography{cite.bib}

\appendix

\section*{Appendix}

\section{Differential Privacy Proof}
\label{appendix-proof}

\subsection{Privacy of Relation Perturbation}
\label{appendix-distance-proof}

\begin{proof}[Proof of Theorem~\ref{thm:relation_mem_privacy}]
Fix finite $\epsilon\geq0$ and $\lambda\geq0$.
For each pair of compared inputs, use the same entity types,
encoder, extended removal distance, and candidate pools with exactly $K$
real predicates, where $K\geq1$ is fixed across compared inputs.
In particular, each $\mathcal C_i$ is independent of the private
predicates and gives the common support
$\mathcal O_i=\mathcal C_i\cup\{\texttt{null}\}$.

\noindent\textbf{Utility sensitivity.}
For any two input relations $r_i,r_i'\in\mathcal R$ compared
under the same candidate pool $\mathcal C_i$, define
\[
\delta_i:=d_{\mathcal R}(r_i,r_i')
=\tfrac12\|z(r_i)-z(r_i')\|_2.
\]
We show that changing the input from $r_i$ to $r_i'$ changes
the utility of any fixed output by at most $\delta_i$.
For a real candidate $c\in\mathcal C_i$, the Euclidean triangle
inequality gives
\[
\begin{aligned}
d_{\mathcal R}(r_i,c)
&=\tfrac12\left\|(z(r_i)-z(r_i'))+(z(r_i')-z(c))\right\|_2\\
&\leq\tfrac12\|z(r_i)-z(r_i')\|_2
     +\tfrac12\|z(r_i')-z(c)\|_2\\
&=\delta_i+d_{\mathcal R}(r_i',c).
\end{aligned}
\]
This step involves only real predicates. Interchanging $r_i$
and $r_i'$ yields the reverse bound, so
\begin{equation}
\begin{aligned}
|u(r_i,c)-u(r_i',c)|
&=|d_{\mathcal R}(r_i,c)-d_{\mathcal R}(r_i',c)|
\leq\delta_i.
\end{aligned}
\label{eq:reverse_triangle}
\end{equation}
For the deletion output, Eq.~\eqref{eq:relation_release_utility}
assigns the same utility $-\lambda$ under both inputs. Hence
\[
|u(r_i,\texttt{null})-u(r_i',\texttt{null})|
=|-\lambda-(-\lambda)|=0\leq\delta_i.
\]
Consequently,
\begin{equation}
|u(r_i,y)-u(r_i',y)|\leq\delta_i,
\qquad y\in\mathcal O_i.
\label{eq:utility_sensitivity}
\end{equation}
\noindent\textbf{Relation privacy.}
Let $\pi_i(y\mid r)$ denote the probability in
Eq.~\eqref{eq:relation_mem}, and write
\[
w_i(r,y)=e^{\epsilon u(r,y)/2},\qquad
Z_i(r)=\sum_{y\in\mathcal O_i}w_i(r,y).
\]
All weights are positive and finite, so $0<Z_i(r)<\infty$.
By Eq.~\eqref{eq:utility_sensitivity},
\begin{equation}
e^{-\epsilon\delta_i/2}w_i(r_i',y)
\leq w_i(r_i,y)
\leq e^{\epsilon\delta_i/2}w_i(r_i',y).
\label{eq:numerator_bound}
\end{equation}
Summing the lower bound over the common output support gives
\begin{equation}
Z_i(r_i)\geq e^{-\epsilon\delta_i/2}Z_i(r_i'),
\qquad
\frac{Z_i(r_i')}{Z_i(r_i)}\leq e^{\epsilon\delta_i/2}.
\label{eq:normalization_bound}
\end{equation}
Combining the weight and normalizer bounds, for every
$y\in\mathcal O_i$ we obtain
\begin{equation}
\begin{aligned}
\frac{\pi_i(y\mid r_i)}{\pi_i(y\mid r_i')}
&=\frac{w_i(r_i,y)}{w_i(r_i',y)}
  \frac{Z_i(r_i')}{Z_i(r_i)}\\
&\leq e^{\epsilon\delta_i/2}e^{\epsilon\delta_i/2}
=e^{\epsilon\delta_i}.
\end{aligned}
\label{eq:single_relation_privacy}
\end{equation}
Summing this inequality over any relation-output event
$\mathcal S\subseteq\mathcal O_i$ establishes
Definition~\ref{def:metric_srdp} for the mechanism sampling $\widetilde r_i$.
\end{proof}

\subsection{Privacy of Document Reconstruction}
\label{appendix-reconstruction-proof}

\begin{proof}[Proof of Theorem~\ref{thm:reconstruction_privacy}]
We first compose the single-relation guarantees, then apply the
postprocessing property to document reconstruction.

\noindent\textbf{Composition across relations.}
Fix $D,D'$ with aligned canonical triples
$\mathcal F=\mathcal F(D)=((s_i,r_i,o_i))_{i=1}^{m}$ and
$\mathcal F'=\mathcal F(D')=((s_i,r_i',o_i))_{i=1}^{m}$.
The endpoints and slot order are fixed, as in the document-level
extension in Section~\ref{sec:metric_srdp}. Let
$\delta_i=d_{\mathcal R}(r_i,r_i')$. For
$\mathbf y\in\prod_i\mathcal O_i$, write
$\mathcal F_{\mathbf y}=((s_i,y_i,o_i))_{i=1}^{m}$.
Placing sampled relations into triples retains the fixed endpoints
and slot order. Since the algorithm samples relations independently,
\begin{equation}
\begin{aligned}
\frac{\Pr[\mathcal M(D)=\mathcal F_{\mathbf y}]}
{\Pr[\mathcal M(D')=\mathcal F_{\mathbf y}]}
&=\prod_{i=1}^{m}
  \frac{\pi_i(y_i\mid r_i)}{\pi_i(y_i\mid r_i')}\\
&\leq e^{\epsilon\sum_i\delta_i}
=e^{\epsilon d_{\mathcal F}(\mathcal F,\mathcal F')}.
\end{aligned}
\label{eq:document_ratio}
\end{equation}
Summing over every $\mathcal F_{\mathbf y}$ in an arbitrary event
$\mathcal S$ proves Eq.~\eqref{eq:document_relation_dp}.
Repeated mentions reuse one canonical triple's sampled outcome
and therefore introduce no additional independent releases.

\noindent\textbf{Released connectivity.}
Keeping the endpoint pair of each non-null output triple is
postprocessing, so the released connectivity view inherits the same
relation-DP bound~\cite{dwork2014algorithmic}.
This guarantee compares relation assignments on fixed, aligned
entity pairs, rather than inputs with different original edge sets.

\noindent\textbf{Text reconstruction.}
Document reconstruction preserves the joint guarantee by
postprocessing~\cite{dwork2014algorithmic}.
Let $\mathcal M_{\mathrm{text}}(D)=
\mathsf{Reconstruct}(\mathcal M(D),A)$.
For the document comparisons in Section~\ref{sec:metric_srdp}, let
$A=A(D)=A(D')$ be the shared masked document. Write
$\mathcal F=\mathcal F(D)$ and $\mathcal F'=\mathcal F(D')$,
and consider a text-output event $\mathcal S$. Define
$K_A(\mathcal S\mid z)=
\Pr[\mathsf{Reconstruct}(z,A)\in\mathcal S]$.
The reconstruction procedure uses the sampled predicates and
placeholders in $A$, without reading original relation values.
Its kernel is therefore identical for the two inputs.
The joint bound in Eq.~\eqref{eq:document_relation_dp} gives
\[
\begin{aligned}
\Pr[\mathcal M_{\mathrm{text}}(D)\in\mathcal S]
&=\sum_z K_A(\mathcal S\mid z)\Pr[\mathcal M(D)=z]\\
&\leq e^{\epsilon d_{\mathcal F}(\mathcal F,\mathcal F')}
  \sum_z K_A(\mathcal S\mid z)\Pr[\mathcal M(D')=z]\\
&=e^{\epsilon d_{\mathcal F}(\mathcal F,\mathcal F')}
  \Pr[\mathcal M_{\mathrm{text}}(D')\in\mathcal S].
\end{aligned}
\]
This proves relation-DP for the reconstructed text under the specified
document comparisons.
\end{proof}

\section{Prompt}
\label{appendix-prompt}
Figure~\ref{fig:relation_extraction_prompt} gives the prompt for
canonical relation extraction.
Figure~\ref{fig:relation_candidate_prompt} generates relation candidates from
an ordered entity-type pair.

\begin{figure*}[!t]
\centering
\captionsetup{font=small,skip=3pt}
\begin{tcblisting}{
    listing only,
    width=\textwidth,
    colback=gray!5,
    colframe=black!60,
    boxrule=0.4pt,
    arc=0pt,
    boxsep=0pt,
    left=1.5mm,
    right=1.5mm,
    top=1mm,
    bottom=1mm,
    before skip=0pt,
    after skip=0pt,
    listing options={
        basicstyle=\ttfamily\fontsize{8}{9}\selectfont,
        breaklines=true,
        breakatwhitespace=true,
        columns=fullflexible,
        keepspaces=true,
        showstringspaces=false,
        aboveskip=0pt,
        belowskip=0pt
    }
}
You are a knowledge graph extraction assistant.
Extract entities and explicitly stated relations from the document, with one numbered sentence per line, and locate every relation occurrence.

1. Entities
- Include only entities belonging to this inventory: PERSON, NORP, FAC, ORG, GPE, LOC, PRODUCT, EVENT, WORK_OF_ART, LAW, LANGUAGE.
- Assign one canonical name per entity; resolve clear pronouns, aliases, and abbreviations. Exclude dates, quantities, titles, professions, abstract concepts, and biological taxa.
- Return canonical names only, without type labels. Entity types are assigned separately from canonical names and the fixed inventory, without source sentences or extracted predicates.

2. Relations
- Extract directed (subject, relation, object) triples only when both endpoints are listed entities.
- Use specific lower-case snake_case predicates, such as works_for or founded. Merge equivalent expressions and output each distinct triple once; retain different relations for the same entity pair.
- Do not infer relations from external knowledge, commonsense, negation, hypotheticals, or speculation.

3. Occurrences
- For every occurrence, return the printed sentence number and the shortest exact substring expressing the relation, usually a verb phrase. Each numbered line contains exactly one sentence.
- Copy the substring character for character. Exclude entity names, dates, and unrelated modifiers; do not return a bare preposition or copula. Skip occurrences without an explicit relation-bearing span.
- Report all occurrences of each triple. Do not report character offsets.

Return only JSON. Triple endpoints must match canonical entity names.
{"entities": [{"name": "<entity_name>"}],
 "triples": [{"subject": "<subject_entity>", "relation": "<canonical_relation>",
   "object": "<object_entity>", "occurrences": [{"sentence": <sentence_number>,
   "predicate": "<exact substring>"}]}]}
Document: {ruled}
\end{tcblisting}
\caption{Prompt for entity and relation extraction.}
\label{fig:relation_extraction_prompt}
\end{figure*}

\begin{figure*}[!t]
\centering
\captionsetup{font=small,skip=3pt}
\begin{tcblisting}{
    listing only,
    width=\textwidth,
    colback=gray!5,
    colframe=black!60,
    boxrule=0.4pt,
    arc=0pt,
    boxsep=0pt,
    left=1.5mm,
    right=1.5mm,
    top=1mm,
    bottom=1mm,
    before skip=0pt,
    after skip=0pt,
    listing options={
        basicstyle=\ttfamily\fontsize{8}{9}\selectfont,
        breaklines=true,
        breakatwhitespace=true,
        columns=fullflexible,
        keepspaces=true,
        showstringspaces=false,
        aboveskip=0pt,
        belowskip=0pt
    }
}
You are a semantic relation calibration assistant.
Given only an ordered entity-type pair and a candidate count K, generate relations from the subject type to the object type. Use no documents, entity identities, or private relation values.

Each candidate must satisfy all three conditions:
1. Type compatibility: The relation must be meaningful for both types in the specified direction. For (ORG, PERSON), employs is valid; works_for has the wrong direction.
2. Non-containment: No candidate may imply another in the pool. For example, do not include both spouse_of and family_member_of; retain the more specific predicate. Check every candidate pair.
3. Semantic diversity: Cover distinct kinds of relationships. Merge synonyms and expressions differing only in tense, voice, or wording. Prefer candidates that convey different information and, where appropriate, are mutually exclusive.

Use concise lower-case snake_case predicates. Avoid vague predicates such as related_to, associated_with, connected_to, involves, part_of, and is_a. Do not name specific entities.
Return exactly {K} distinct candidates satisfying all three conditions.

Input:
Subject Type: {subject_type}
Object Type: {object_type}
Candidate Count: {K}
Return only JSON:
{"relation_pool": ["<relation_1>", "<relation_2>", "..."]}
\end{tcblisting}
\caption{Prompt for type-conditioned candidate generation.}
\label{fig:relation_candidate_prompt}
\end{figure*}

\section{Additional Victim Model: Gemini-2.5-Flash}
\label{appendix-gemini-results}

\begin{table*}[t]
\centering
\caption{Privacy--utility trade-off of \system\ with Gemini-2.5-Flash.}
\label{tab:privacy_utility_budgets_gemini}
\small
\setlength{\tabcolsep}{4pt}
\renewcommand{\arraystretch}{0.7}
\resizebox{\textwidth}{!}{
\begin{tabular}{ll*{10}{r}}
\toprule
\multirow{3}{*}{\textbf{Victim}} &
\multirow{3}{*}{\textbf{Config}} &
\multicolumn{5}{c}{\textbf{2WikiMultiHopQA}} &
\multicolumn{5}{c}{\textbf{HotpotQA}} \\
\cmidrule(lr){3-7}\cmidrule(lr){8-12}
& & \multicolumn{2}{c}{GRASP} & \multicolumn{2}{c}{GraphSteal}
& \multirow{2}{*}{QA F1 $\uparrow$}
& \multicolumn{2}{c}{GRASP} & \multicolumn{2}{c}{GraphSteal}
& \multirow{2}{*}{QA F1 $\uparrow$} \\
\cmidrule(lr){3-4}\cmidrule(lr){5-6}\cmidrule(lr){8-9}\cmidrule(lr){10-11}
& & Conn. F1 $\downarrow$ & Rel. F1 $\downarrow$
& Conn. F1 $\downarrow$ & Rel. F1 $\downarrow$ &
& Conn. F1 $\downarrow$ & Rel. F1 $\downarrow$
& Conn. F1 $\downarrow$ & Rel. F1 $\downarrow$ & \\
\midrule
\multirow{5}{*}{HippoRAG 2} & no-DP & 76.83 & 71.50 & 85.43 & 77.71 & 65.10 & 75.95 & 68.51 & 72.72 & 61.95 & 75.62 \\
 & $\epsilon=10$ & 76.61 & 18.52 & 75.89 & 19.73 & 57.53 & 70.71 & 23.47 & 71.23 & 28.11 & 70.03 \\
 & $\epsilon=5$ & 74.59 & 13.21 & 73.80 & 16.74 & 55.31 & 66.13 & 16.95 & 63.19 & 19.68 & 64.06 \\
 & $\epsilon=1$ & 55.43 & 11.53 & 56.50 & 14.98 & 53.62 & 61.34 & 16.71 & 61.54 & 18.87 & 66.63 \\
 & $\epsilon=0.1$ & 55.14 & 10.14 & 55.66 & 14.31 & 53.01 & 56.40 & 13.69 & 60.69 & 17.18 & 66.27 \\
\midrule
\multirow{5}{*}{LightRAG} & no-DP & 66.18 & 59.19 & 79.13 & 74.23 & 62.48 & 75.19 & 69.55 & 80.11 & 66.81 & 70.19 \\
 & $\epsilon=10$ & 61.49 & 17.67 & 76.11 & 33.04 & 61.43 & 64.00 & 26.22 & 77.53 & 44.50 & 68.83 \\
 & $\epsilon=5$ & 57.11 & 13.55 & 74.99 & 28.83 & 58.89 & 62.77 & 26.05 & 77.37 & 41.47 & 65.60 \\
 & $\epsilon=1$ & 47.19 & 12.19 & 70.12 & 22.35 & 58.22 & 61.92 & 22.27 & 75.99 & 35.74 & 65.48 \\
 & $\epsilon=0.1$ & 41.18 & 10.09 & 63.55 & 20.44 & 58.11 & 60.56 & 21.04 & 74.16 & 30.74 & 64.96 \\
\midrule
\multirow{5}{*}{PathRAG} & no-DP & 73.50 & 67.44 & 71.41 & 58.77 & 67.03 & 74.50 & 64.48 & 69.97 & 54.59 & 68.87 \\
 & $\epsilon=10$ & 72.89 & 19.16 & 71.93 & 36.91 & 64.34 & 72.01 & 28.25 & 68.64 & 36.47 & 66.92 \\
 & $\epsilon=5$ & 68.79 & 12.80 & 63.97 & 32.53 & 61.57 & 67.48 & 26.46 & 65.46 & 36.12 & 66.01 \\
 & $\epsilon=1$ & 52.51 & 12.14 & 54.81 & 28.03 & 54.23 & 61.99 & 20.66 & 57.39 & 32.76 & 65.36 \\
 & $\epsilon=0.1$ & 52.17 & 7.10 & 50.84 & 22.30 & 51.30 & 54.12 & 20.06 & 50.63 & 32.08 & 65.27 \\
\bottomrule
\end{tabular}
}
\end{table*}

Table~\ref{tab:privacy_utility_budgets_gemini} examines the privacy--utility
trade-off for $\epsilon\in\{10,5,1,0.1\}$ across both datasets and
all three GraphRAG systems, using Gemini-2.5-Flash as the victim generator.
No-DP uses the original document; the other rows use documents
sanitized by \system\ at the indicated budget. All scores are
percentages. Overall, smaller $\epsilon$ reduces connection and
relation recovery at the cost of QA utility, while larger $\epsilon$
retains more utility but permits more recovery.

For LightRAG on 2WikiMultiHopQA, decreasing $\epsilon$ from 10 to 1
and then to 0.1 reduces GRASP Conn. F1 from 61.49 to 47.19 and
41.18, and Rel. F1 from 17.67 to 12.19 and 10.09. GraphSteal shows
the same trend: Conn. F1 decreases from 76.11 to 70.12 and 63.55,
while Rel. F1 decreases from 33.04 to 22.35 and 20.44.
QA F1 correspondingly declines from 61.43 to 58.22 and 58.11.
For PathRAG on HotpotQA, reducing $\epsilon$ from 10 to 0.1 lowers
GRASP Conn. F1 from 72.01 to 54.12 and Rel. F1 from 28.25 to
20.06, while QA F1 falls from 66.92 to 65.27. The utility cost
varies across systems and datasets: on 2WikiMultiHopQA, the same
budget reduction lowers PathRAG's QA F1 from 64.34 to 51.30.
Although individual scores are not uniformly monotone, the overall
trend is consistent with Section~\ref{sec:rq2}: smaller privacy
budgets provide stronger privacy protection with lower QA utility,
and larger budgets exhibit the reverse trade-off.

\section{Attack Details}
\label{appendix-attack-details}

We use GRASP and targeted GraphSteal to evaluate the two attack
objectives in Section~\ref{sec:threat_model}: connectivity inference,
which identifies entities directly connected to the target, and
relation semantics recovery, which identifies how these entities
are related to the target.

\noindent\textbf{GRASP.}
GRASP~\cite{song2026graphs} anchors retrieval on the target entity and
frames extraction as a structured relation-processing task, requesting
records with an instance identifier, source entity, relation predicate,
and destination entity. The endpoints support inference of the target's
one-hop connections, while the predicates support recovery of relation
semantics. Instance identifiers associate fields within each record
to reduce cross-record mixing. We preserve relation direction and
restrict predictions to target-incident relations. New predictions are identified against the
accumulated triples; instance identifiers are excluded from scoring.

\noindent\textbf{Targeted GraphSteal.}
Targeted GraphSteal~\cite{gu2026graphsteal} starts with a known target
entity and uses entities and relations discovered in previous answers
to guide subsequent queries, progressively recovering the target's
local relations. Its outputs include both neighboring entities and
relation descriptions: the former support connectivity inference,
while the latter support recovery of relation semantics. We parse
answers into directed subject--predicate--object triples, retain
target-incident predictions, and merge repeated results across turns
before computing Conn. and Rel. scores.

\section{Privacy Metric Details}
\label{appendix-privacy-metrics}

\emph{Connection Recovery (Conn.)} checks whether the attacker
identifies the correct directed entity pair, regardless of the
predicted predicate. Entity aliases are mapped to canonical names
before comparison. \emph{Relation Recovery (Rel.)} additionally
checks whether the predicted predicate captures the true relation's
meaning. GPT-4o-mini accepts semantically matching predicates
or broader predictions entailed by the true relation. For example,
if Alice is Bob's spouse, predicting that Alice is Bob's family
member counts as relation recovery; predicting a spouse relation
from a true family-member relation does not. Predictions without
predicates are evaluated only for connection recovery.

Precision measures the fraction of predictions that are correct;
recall measures the fraction of true items recovered. For target
$\ts$, let $E_{\ts}$ and $\widehat E_{\ts}$ denote the true and
predicted triple sets. For relation recovery, we use these triples
directly: $E_{\ts}^{\mathrm{Rel}}=E_{\ts}$. For connection recovery,
we discard predicates and keep distinct directed pairs:
$E_{\ts}^{\mathrm{Conn}}=\{(u,v):(u,r,v)\in E_{\ts}\}$.
Predictions are transformed in the same way.
For $a\in\{\mathrm{Conn},\mathrm{Rel}\}$, let $h_{\ts}^a$ be the
number of correct matches under the corresponding rule. We compute
\begin{equation}
p_{\ts}^a=\frac{h_{\ts}^a}{\max(1,|\widehat E_{\ts}^a|)},
\quad
r_{\ts}^a=\frac{h_{\ts}^a}{|E_{\ts}^a|}.
\label{eq:recovery_precision_recall}
\end{equation}
All evaluated targets have nonempty true sets. Empty predictions
receive zero precision and recall.

We average target-level precision and recall over the target set $T$.
For connection recovery,
\begin{equation}
P_{\mathrm{Conn}}=\frac{\sum_{\ts\in T}p_{\ts}^{\mathrm{Conn}}}{|T|},
\quad
R_{\mathrm{Conn}}=\frac{\sum_{\ts\in T}r_{\ts}^{\mathrm{Conn}}}{|T|}.
\label{eq:reconstruction_macro_scores}
\end{equation}
For relation recovery,
\begin{equation}
P_{\mathrm{Rel}}=\frac{\sum_{\ts\in T}p_{\ts}^{\mathrm{Rel}}}{|T|},
\quad
R_{\mathrm{Rel}}=\frac{\sum_{\ts\in T}r_{\ts}^{\mathrm{Rel}}}{|T|}.
\label{eq:relation_macro_scores}
\end{equation}
Conn. F1 and Rel. F1 are the harmonic means of their respective
averaged precision and recall, computed as
$F1_a=2P_aR_a/(P_a+R_a)$ for $a\in\{\mathrm{Conn},\mathrm{Rel}\}$.

\section{Defense Baseline Details}
\label{appendix-defense-baselines}

\looseness=-1
\noindent\textbf{Abstractive.}
Abstractive~\cite{liu2026exposing} instructs the generator to select
query-relevant evidence and express it concisely in its own words.
We use the system-prompt implementation evaluated by
Song et al.~\cite{song2026graphs}. The instruction supplements the
base prompt and returns \texttt{NO\_RELEVANT\_CONTENT} when the
context is irrelevant.

\looseness=-1
\noindent\textbf{ID Alignment.}
ID Alignment~\cite{song2026graphs} assigns the same displayed identifier
(e.g., 1) to every retrieved relation record before passing the context
to the generator. Endpoints, relation descriptions, and row order are
preserved. Collapsing identifiers removes the record-level distinction
used by structured extraction and can induce cross-record mixing.
The transformation applies during context serialization, leaving the
stored graph unchanged.

\looseness=-1
\noindent\textbf{Decoy.}
Decoy~\cite{song2026graphs} adds misleading \texttt{src\_entity} and
\texttt{dst\_entity} columns to retrieved relation records while retaining
the true \texttt{src} and \texttt{dst} fields. The added values differ
from the true endpoints. A system instruction identifies the added
fields as internal metadata to ignore in answers. This presentation
can steer extraction toward incorrect endpoint assignments while
preserving the true evidence for QA. Decoy is evaluated separately
from ID Alignment.

\looseness=-1
\noindent\textbf{Protective Prompt.}
Protective Prompt implements the Protective System Prompt mitigation
in GraphSteal~\cite{gu2026graphsteal}. It prepends a protective
instruction to the generator's system prompt, directing the LLM not
to disclose content retrieved from the knowledge base. The source
corpus, stored graph, and retrieved evidence remain unchanged.
The instruction applies when answering both benign and attack queries;
attack predictions are extracted from the resulting answers.

\end{document}